\documentclass[prl, aps, amsmath, amssymb, twocolumn, superscriptaddress,nofootinbib]{revtex4-1}

\usepackage{graphicx}
\usepackage{hyperref}
\usepackage{physics}
\usepackage{color}
\usepackage{enumerate}
\usepackage{bm,ulem}
 \usepackage[caption=false]{subfig}

\begin{document}

\title{Geometric phase through spatial potential engineering}

\author{Stefano Cusumano}
\email[Corresponding author: ]{stefano.cusumano@sns.it}
\affiliation{NEST, Scuola Normale Superiore and Istituto Nanoscienze-CNR, I-56127 Pisa, Italy}

\author{Antonella De Pasquale}
\affiliation{NEST, Scuola Normale Superiore and Istituto Nanoscienze-CNR, I-56127 Pisa, Italy}
\affiliation{Dipartimento di Fisica e Astronomia, Universit\'a di Firenze, I-50019, Sesto Fiorentino (FI), Italy}
\affiliation{INFN Sezione di Firenze, via G.Sansone 1, I-50019 Sesto Fiorentino (FI), Italy}

\author{Vittorio Giovannetti}
\affiliation{NEST, Scuola Normale Superiore and Istituto Nanoscienze-CNR, I-56127 Pisa, Italy}

\begin{abstract}
We propose a spatial analog of 
the Berry's phase mechanism for the coherent manipulation of  states of non-relativistic massive particles moving  in a two-dimensional landscape.
In our construction the temporal modulation of the system Hamiltonian is replaced by a modulation of the confining potential along
the transverse direction of the particle propagation. 
By properly tuning the model parameters the resulting scattering input-output relations exhibit 
a  Wilczek-Zee non-abelian phase shift contribution that is intrinsically geometrical,
 hence insensitive to the specific details of the potential landscape. 
A theoretical derivation of the effect is provided together with practical examples.
\end{abstract}

\maketitle

In recent years a  strong demand for  developing quantum engineering~\cite{qse_review,qse_optics,loyd} procedures
has been fostered by the huge technological development requiring faster and more efficient circuits and transistors, but also by the first prototypes of quantum computers. 
As the main resource for quantum supremacy ultimately relay on the amount of quantum coherence
one can store on a system,  the ability 
to design control schemes that allow for its manipulation  
becomes of paramount importance~\cite{rossini}. 
A possibility in this direction is presented by applications of 
the non-abelian generalization~\cite{WIL} of the Berry phase mechanism~\cite{berry}.
In these approaches~\cite{ZANA,HOL1,HOL2,HOL2,HOL3,HOL4,HOL5,HOL6,dechiara,yale,berry_experimental,hansom} a target (state independent) 
transformation is implemented by
driving the system Hamiltonian
along a closed path in the control
parameter space either adiabatically as in the original proposal~\cite{berry}, or nonadiabatically~\cite{NONAD}. 
The resulting operation (typically referred to as {\it holonomy}) has an intrinsic 
geometrical character~\cite{carollo1,carollo2} that makes it resilient to 
local fluctuations~\cite{NOIS1,NOIS2}, 
hence offering an attractive alternative 
to quantum error-correction techniques~\cite{QEC,dalibard}. 

\begin{figure}[!t]
\centering
\includegraphics[scale=0.4]{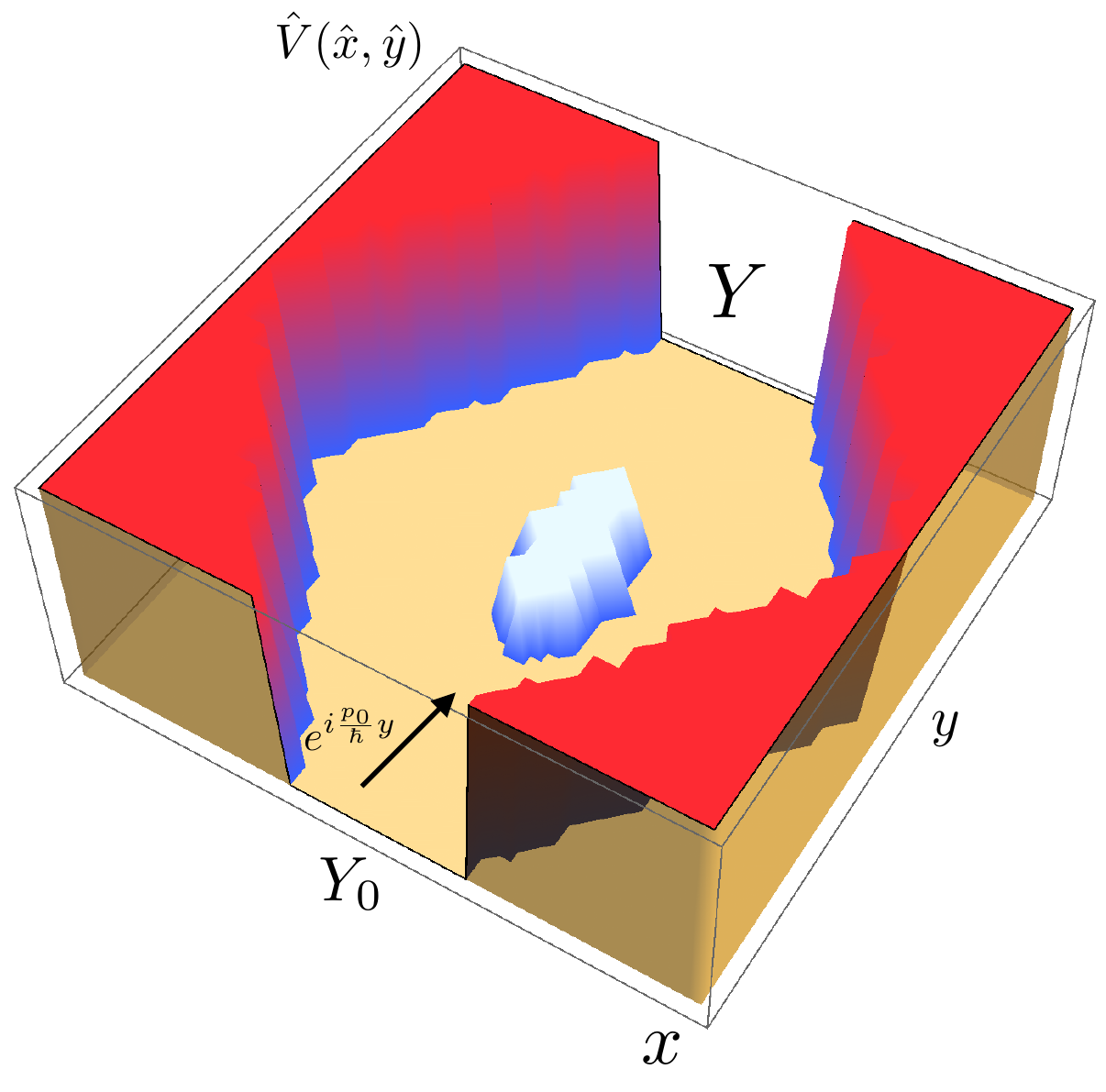}
\caption{(Color online) Pictorial  representation of a 2D potential landscape. As the particle $A$ is moving forward along the longitudinal $y$ axis, it sees a varying potential along the transverse axis $x$ defined by the scattering potential 
${V}(\hat{x},\hat{y})$.}
\label{fig:potential_example}
\end{figure}

Inspired by the above approaches we present here a proposal for the coherent
manipulation of  a 
non-relativistic massive particle $A$ through holonomies obtained by properly engineering the potential landscape it
experiences when traveling through a scattering region.
Although the scheme can be in principle applied to arbitrary spatial configurations, we shall focus on 2D geometries (see Fig.~\ref{fig:potential_example})  where  
desired potential profiles with a high degree of accuracy and low numbers of impurities~
can be easily achieved in semiconductor platforms, either by 
 direct nanofabrication~\cite{esaki_superlattice,cardona,park,lee,newref}, or via external 
gate potential techniques. 
As a further simplification,  the kinetic energy of the incoming particle will be taken to 
be the largest of the model. While not being essential,  this assumption 
 allows us to isolate in the solution of the Schr\"odinger equation
 the geometric term (the holonomy) from an irrelevant dynamical phase.

\paragraph{The model:--}
Let $A$ be a non-relativistic particle  of mass $m$, propagating in the $xy$-plane, 
under the action of a scattering potential ${V}(\hat{x},\hat{y})$, so that the resulting Hamiltonian is $\hat{H}:=\frac{\hat{p}_x^2}{2m}+\frac{\hat{p}_y^2}{2m}+{V}(\hat{x},\hat{y})$.
As shown in Fig.~\ref{fig:potential_example} we assume $A$ to enter the setup with assigned energy $E$ 
corresponding to an input state 
that, far away from the scattering center in the negative $y$ direction,  
 is described by an impinging plane-wave with assigned momentum $p_0>0$ which sets the largest energy scale in the system
 (i.e. $E_{\text{kin}}:= p_0^2/(2m) \simeq E$): 
adopting the scattering formalism we analyze the dynamics of  the particle by looking for solutions of the time-independent Schr\"odinger equation $\hat{H}\ket{\psi_E}=E\ket{\psi_E}$
 that are compatible with the chosen boundary conditions. 
Moving to a representation with respect to the $y$ coordinate, we hence cast this equation in the form
\begin{eqnarray}
\label{eq:schrodinger}
-\tfrac{\hbar^2}{2m}\partial^2_y|{\psi}_E(y)\rangle+\hat{h}_y(\hat{x})\ket{\psi_E(y)}=E\ket{\psi_E(y)}\;,
\end{eqnarray}
with $\ket{\psi_E(y)}:=\bra{y}\ket{\psi_E}$ being the transverse wavevector component for 
fixed  longitudinal position 
($\ket{y}$ being the eigenstate of the of the position operator $\hat{y}$).
In Eq.~(\ref{eq:schrodinger}) the self-adjoint operator 
$\hat{h}_y(\hat{x}):=\frac{\hat{p}^2_x}{2m}+{V}_y (\hat{x})$
is  the transverse Hamiltonian where ${V}_y(\hat{x}):={V}(\hat{x},y)$ 
is obtained by   replacing in ${V}(\hat{x},\hat{y})$ the operator $\hat{y}$  with 
its  eigenvalue $y$.  Therefore, from now on $y$ will be treated as a real variable, while $\hat x$ is still an operator. Without loss of generality in what follows we shall assume the parametric dependence upon $y$ of ${V}_y(\hat{x})$ to be mediated via a collection of (real) {\it control} functions, 
which we represent collectively as components of the  vector $\vec{R}_y:=(R_y^{(1)}, R_y^{(2)},\cdots)$, i.e.
\begin{eqnarray}\label{DEPY} 
{V}_y(\hat{x}) = {V}_0(\hat{x};{\vec{R}_y})\;.
\end{eqnarray} 
For the sake of simplicity, we shall then
assume $\hat{h}_y(\hat{x})$ to have discrete spectrum, for instance forcing the potential ${V}_y(\hat{x})$  to induce local transverse confinement
 for all values of $y$. Thus, for fixed $y$, we identify 
 the eigenvectors of $\hat{h}_y(\hat{x})$ with the discrete orthonormal set $\{ |{\phi_y^{(\ell)}}\rangle ; {\ell =0,1,2,\cdots}\}$, the associated
 eigenvalues being the quantities $E_y^{(\ell)}:= \tfrac{\hbar^2  }{2m}\epsilon_y^{(\ell)}$, which we  assume to be in increasing order with respect to the index $\ell$. 
Decomposing hence $\ket{\psi_E(y)}$ as 
$\ket{\psi_E(y)}=\sum_\ell C_y^{(\ell)}|{\phi_y^{(\ell)}}\rangle$
with $C_y^{(\ell)}$ being complex amplitudes,  and introducing the rescaled energy $\epsilon:= \tfrac{2m }{\hbar^2}E$, without
any approximations , as shown in Sec. II of Supplemental Material (SM) we can recast Eq.~\eqref{eq:schrodinger} as:
\begin{equation}
 \label{eq:differential_1}
\left(\partial_y +  K_y\right)^2 {{\bf C}_y}     
+ (  \epsilon -  \Omega_y 
){{\bf C}_y}  =0   \;,
\end{equation} 
where ${\bf C}_y$ is the column vector of components $(C_y^{(0)}, C_y^{(1)},C_y^{(2)},\cdots)$, 
 and $\Omega_y$ is an Hermitian matrix with elements  $[\Omega_y]_{\ell \ell'}  := \epsilon_y^{(\ell)} \delta_{\ell\ell'}$,
$\delta_{\ell\ell'}$ being the Kronecker delta. In the above expression 
 $K_y = - K_y^\dag$ is a real anti-Hermitian matrix which ultimately triggers the coupling  among the various components of ${\bf C}_y$ with an intensity that scales with the inverse 
of the gaps of the associated local  energies $E_y^{(\ell)}$, i.e. 
\begin{equation} \label{IMPOKY} 
{[K_y]}_{\ell \ell'} = \langle{\phi_{y}^{(\ell)}}|{\partial_y{\phi}_y^{(\ell')}}\rangle=\frac{\langle{\phi_y^{(\ell)}}|\big( \partial_y{{V}}_y(\hat{x})\big) |{\phi_y^{(\ell')}}\rangle}{E_y^{(\ell')}-E_y^{(\ell)}} (1- \delta_{\ell\ell'})\;, 
\end{equation} 
see Secs. I and II of the SM for details.
As in  Refs.~\cite{ALDINGER,GROSSO,ALDEN,levy}  the presence of $K_y$ 
can be thought as arising via  minimal coupling from a non-abelian vector potential: accordingly it  can  be
 gauged away 
 through the action of the unitary mapping 
induced by the path-ordered exponential  ${\cal U}_{Y_0\rightarrow y}:=P
{\exp}\left[-\int^y_{{Y_0}} {K}_{y'}dy'\right]$, ${Y_0}$ being the longitudinal coordinate  defining the beginning of the scattering region. 
Specifically by setting 
${{\bf C}_y}  = {\cal U}_{Y_0\rightarrow y} \;  \tilde{{\bf C}}_y$
we can rewrite~(\ref{eq:differential_1}) as the following spinor 1D Schr\"{o}dinger equation 
\begin{equation}\partial_y ^2 \tilde{\bf C}_y   + ( \epsilon -  \tilde{\Omega}_y )\tilde{\bf C}_y  =0
 \label{eq:differential_2}
    \;,
\end{equation} 
with $\tilde{\Omega}_y :=   {\cal U}_{Y_0\rightarrow y}^\dag {\Omega}_y
   {\cal U}_{Y_0\rightarrow y}$ holding the same spectrum of $\Omega_y$ and
playing the role of an effective potential. For sufficiently smooth potential modulations and assuming 
~$\epsilon$ (hence the rescaled kinetic component of the incoming particle  $E_{\text{kin}}$) to be the largest energy scale in the system, Eq.~(\ref{eq:differential_2}) admits solutions which, according to the Wentzel-Kramers-Brillouin (WKB) approximation method~\cite{MESSIAH}, read
 $\tilde{\bf{C}}_{y}= {\cal W}^{(+)}_{y}   {\bf{A}} +  {\cal W}^{(-)}_{y}   {\bf{B}}$,
with the vectors ${\bf{A}}$, ${\bf{B}}$ being determined by the boundary conditions of the problem and with the matrices  ${\cal W}^{(\pm)}_{y}$ describing respectively the
left-to-right and right-to-left propagations of the particle in the sample -- see Sec. III of the SM for details.
In particular  in the very large $\epsilon$ limit, i.e $\epsilon\gg \epsilon_y^{(\ell)}$ for all $y$ and for all the energy levels $ \epsilon_y^{(\ell)}$ involved in the process, we can safely conclude that all $ \epsilon_y^{(\ell)}$ will yield approximately the same phase whose leading contribution can be expressed as ${\cal W}_{y}^{(\pm)}  \simeq e^{ \pm i \sqrt{\epsilon} (y- Y_0)}$.
 In other words, ${\cal W}_{y}^{(\pm)}$ will
 explicitly depend upon the  length of the integration domain, and as it will be clarified in the next section,  contribute to the final solution with an irrelevant global phase [see \eqref{SOLF}].

{\it Holonomy:--} 
Consider now the case  where the particle $A$  propagates from left-to-right  
in a scattering region located in the spatial domain ${\cal I}:= [Y_0,Y]$. 
Setting 
 ${\bf{A}}={\bf{C}}_{Y_0}$, ${\bf{B}}=0$ we can then express
the solution of Eq.~(\ref{eq:differential_1}) at $y=Y$ as
 \begin{equation}\label{SOLF}
{\bf{C}}_Y = 
{\cal U}_{Y_0 \rightarrow Y}   {\cal W}^{(+)}_{Y}  {\bf{C}}_{Y_0}  \;,
\end{equation} 
which, excluding the presence of the counter-propagating contribution ${\cal W}^{(-)}_{Y}$, formally accounts on neglecting reflection effects induced by the scattering region (a regime we can always achieve for large enough values of ${\epsilon}$).  
 The term ${\cal U}_{Y_0\rightarrow Y}$ 
has a purely  holonomic character, introducing   a geometrical non-abelian phase shift in the model. 
To see this explicitly 
notice that  from
Eq.~(\ref{DEPY}) it follows that  the $y$-functional dependence of the vectors $|{\phi_y^{(\ell)}}\rangle$ is fully mediated by the
vector $\vec{R}_y$, i.e. 
\begin{eqnarray} |{\phi_y^{(\ell)}}\rangle = |{\phi_{{\vec{R}}_y}^{(\ell)}}\rangle\;. \label{equiv} 
\end{eqnarray}  Hence the matrix  ${K}_y$ can be equivalently expressed as
\begin{equation} \label{DEFKK} 
{K}_y=  \vec{{K}}(\vec{R}_y)\cdot \partial_y{\vec{R}}_y\;,
\quad  
[\vec{{K}}(\vec{R})]_{\ell\ell'}:=\langle{\phi_{\vec{R}}^{(\ell)}}| \vec{\nabla}_{\vec{R}}\phi_{\vec{R}}^{(\ell')}\rangle,
\end{equation} 
which formally represents the Berry connection of the model~\cite{berry}.
Assume hence  that the trajectory ${\cal R}:= \{\vec{R}_{y} \}_{y\in [Y_0,Y]}$ followed by the  vector $\vec{R}_{y}$ in the
control parameters space 
forms a closed curve (i.e. $\vec{R}_{Y}= \vec{R}_{{Y_0}}$).
 We  can then use Eq.~(\ref{DEFKK}) to 
write 
 ${\cal U}_{Y_0 \rightarrow Y}$ 
 as a path-ordered integral of  the vector field $\vec{{K}}(\vec{R})$  
 along~${\cal R}$, i.e. 
  \begin{eqnarray}
\label{eq:u_gamma}
{\cal U}_{Y_0 \rightarrow Y}  &=& {\cal U}({\cal R}):= P \exp[ - \oint_{\cal R}  d \vec{R} \cdot \vec{{K}}(\vec{R})] \;, 
\end{eqnarray} 
which no longer  depends upon the ``speed" $\partial_y \vec{R}$ of the longitudinal variation of the potential,
making manifest the geometrical nature of the resulting operation. 
Notice  that invoking the non-abelian version of the Stokes theorem~\cite{HALPEN}, the above expression can also be cast into a surface integral associated with the curvature tensor of  $\vec{{K}}(\vec{R})$, see~\cite{STOKES1} and references therein. 
The resulting formula, while being more evocative, is possibly less informative and we report it only in Sec. IV of the SM.

\paragraph{Two-dimensional models:--} 
To be more quantitive
we now focus on the case 
where  the dynamics can be reduced to a two folds Hilbert subspace spanned, say, only by the eigenstates $|\phi_y^{(0)}\rangle$,  $|\phi_y^{(1)}\rangle$
of the transverse Hamiltonian $\hat{h}_y(\hat{x})$. From Eqs. \eqref{eq:differential_1} and \eqref{IMPOKY} this is guaranteed provided that two conditions are satisfied: $i)$ $\Delta_y:= \epsilon_y^{(1)} -\epsilon_y^{(0)}$ is the smallest energy gap for all $y$, such that $[K_y]_{\ell \ell'}$ results negligible for all the other choices of $\ell, \ell'$; $ii)$ ${V}_y(\hat{x})$ is a sufficiently slowly varying function of $y$ so as to avoid unwanted couplings with other energy levels (a condition which is in agreement with the WKB approximation we
already assumed).
Accordingly we can now write $\Omega_y : = \omega_y I - \Delta_y \sigma_3/2$,
where $I$ is the $2\times 2$ identity matrix while  ${\sigma}_3:={\tiny \left[ \begin{array}{cc}
1 & 0  \\ 
0 & -1\end{array} \right]}$, and where
$\omega_y : = (\epsilon_y^{(1)} + \epsilon_y^{(0)})/2$.
Most importantly, ${K}_y$ reduces to a $2\times 2$ matrix proportional to the second Pauli matrix  ${\sigma}_2:=\tiny{\left[ \begin{array}{cc}
0 &  -i  \\ 
i & 0\end{array} \right]}$, i.e. 
\begin{eqnarray}  \label{MATRIX2} 
K_y= i  \lambda_y \sigma_2 \;, \qquad   \lambda_y = \lambda_y^*:= \langle \phi_y^{(0)}| \partial_y \phi_y^{(1)}\rangle\;,
\end{eqnarray} 
which produces trivial auto-commutators   $[{K}_y,{K}_{y'}]=0$ for all  $y,y'$. Accordingly the expression for ${\cal U}_{Y_0\rightarrow y}$ simplifies to
the following SU(2) rotation
${\cal U}_{Y_0\rightarrow y}=\exp[{-i\alpha_{y} {{\sigma}_2} }]$, where $\alpha_{y}:= \int_{Y_0}^y dy' \lambda_{y'}$. 
In particular, for $y=Y$, this allows us to write (\ref{eq:u_gamma}) as 
\begin{eqnarray} \label{HOLONOMY1} {\cal U}({\cal R})=e^{-i\alpha {{\sigma}_2} },\end{eqnarray}  with
\begin{eqnarray}
 \alpha := \oint_{\cal R}
d\vec{R} \cdot \vec{\lambda}(\vec{R}) =  \int_{\cal S}
d\vec{S} \cdot \Big(\vec{\nabla}_{\vec{R}} \times \vec{\lambda}(\vec{R})\Big)\;,  \label{DEFA} 
\end{eqnarray} 
where exploiting~(\ref{equiv}) we write $\vec{\lambda}(\vec{R}):=
\langle{\phi_{\vec{R}}^{(\ell_0)}}|\vec{\nabla}_{\vec{R}}|\phi_{\vec{R}}^{(\ell_1)}\rangle$, and where  in the second identity, following from the standard  (abelian) version  of the Stokes theorem~\cite{STAND,STOKES1}, the integral is performed on 
a regular surface ${\cal S}$ of the control parameter space which admits ${\cal R}$ as bounding curve.
\begin{figure}[!t]
\centering
\includegraphics[scale=0.3]{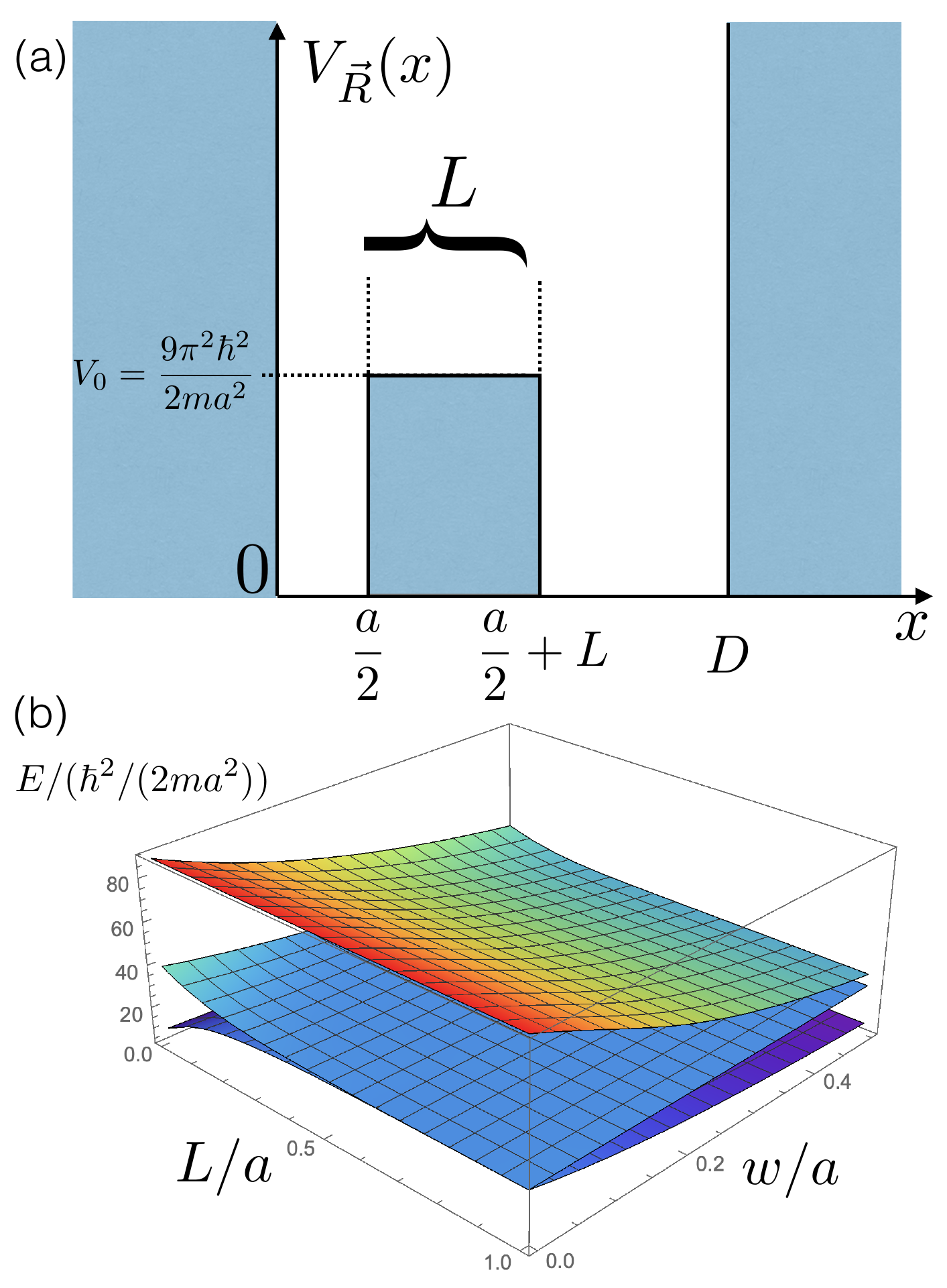}
\caption{(a) Sketch of the structured infinite well potential $V_y(x)= V_{\vec{R}_y}(x)$ discussed in the text.
The model is characterized by a two-dimensional control vector $\vec{R}_y= (L(y), w(y))$
with 
the positive quantities $L(y)$ and $w(y)$ carring the $y$-dependence.
For assigned $\vec{R}$, $V_{\vec{R}}(x)$  exhibit a potential step of constant height $V_0= 9 \pi^2 \hbar^2/(2 m a^2)$ and  width $L$ located 
 at fixed  distance $a/2$ from
the left border of an infinite well that contains it. The total length of the infinite well is also variable and equal to 
 $D= a + L+w$. 
(b) 
Plot of the first three  energy levels $E_0$, $E_1$ and $E_2$ of the model (see Eq.~(\ref{eq:schrodinger}))  as a function of $L$ and $w$ for fixed $a$. 
Here the  energies are measured in unit of $\hbar^2/(2m a^2)$ while the controls in units of~$a$.
  }
\label{fig:energies}
\end{figure}
Inserting this  into Eq.~(\ref{eq:differential_2}) finally gives 
\begin{equation}
\partial_y ^2 \tilde{\bf C}_y   + [   ( \epsilon  - \omega_y)  I + \Delta_y \tilde{\sigma}_3/2]\tilde{\bf C}_y  =0
    \;,
\end{equation} 
with 
$\tilde{\sigma}_3 =e^{i\alpha_{y} {{\sigma}_2} } \sigma_3 e^{-i\alpha_{y} {{\sigma}_2} }$.
Assuming now, as for the general case discussed in the previous section, $\epsilon$ to be the largest energy scale in the system,  i.e. imposing
 $\epsilon \gg |\omega_y |\;, |\Delta_y|$,
the above equation can be integrated under {WKB} approximation yielding
${\cal W}_{Y}^{(\pm)}  \simeq e^{\pm i \int_{Y_0}^Y \sqrt{\epsilon  - \omega_{y'}} dy'}$
 which, although constituting  a refinement  of the solution   ${\cal W}_{Y}^{(\pm)}  \simeq e^{ \pm i \sqrt{\epsilon} (Y- Y_0)}$, still acts on ${\bf{C}}_{Y_0}$ as an irrelevant global phase shift.
Accordingly from~(\ref{SOLF}) we can conclude that  when emerging from the scattering region the transverse component of the wave function of $A$ gets modified via the holonomic rotation  (\ref{HOLONOMY1}), resulting in the following one-qubit gate transformation 
\begin{eqnarray} 
&&|\psi_E(Y_0)\rangle = 
a |\phi_y^{(0)}\rangle + b |\phi_y^{(1)}\rangle\rightarrow  \nonumber \\
&&\quad \rightarrow
|\psi_E(Y)\rangle \simeq  e^{ i \int_{Y_0}^Y \sqrt{\epsilon  - \omega_{y'}} dy'}
 \Big[ (a \cos\alpha - b \sin\alpha) |\phi_y^{(0)}\rangle \nonumber \\ \label{eq:holonomic_gate}
&&\qquad \qquad \qquad + (b \cos\alpha  + a \sin\alpha) |\phi_y^{(1)}\rangle\Big]\;, 
\end{eqnarray} 
 $a$ and $b$ being arbitrary complex amplitudes.
As a final remark notice that,
  since $\alpha$ does not bear any functional dependence upon the input energy $\epsilon$,  
  the effect  can be
 easily generalized to the cases where 
 the longitudinal component of the incoming wave function of $A$ is a wave packet given by the superposition of plane waves involving with different kinetic energies, 
 as long as the latter are much larger than the energy gap between the two levels on which the holonomy acts.
 
As already observed, having $[{K}_y,{K}_{y'}]=0$ greatly simplifies the calculations. The drawback is that under this condition
all the generated holonomy will commute hence allowing us only to span an abelian subgroup of all possible unitaries of the system. 
As discussed explicitly in Sec. V of the SM this limitation however can be overcome by concatenating in series different modulation regions where 
the spatial potential selectively couple  different pairs of energy levels, e.g. first
 $\epsilon_y^{(0)}$ and $\epsilon_y^{(1)}$ then $\epsilon_y^{(0)}$ and $\epsilon_y^{(2)}$ etc., introducing hence extra generators for the holonomy which do not commute. 
\paragraph{Example:--} To test the construction described in the previous section we consider  the case of a  structured infinite potential  well $V_{y}(x)=V_{\vec{R}_y}(x)$  characterized by 
a two dimensional control vector $\vec{R}_y$ with cartesian components 
$R^{(1)}_y := L(y)$ and $R^{(2)}_y:=w(y)$ associated with two positive spatial parameters. Specifically we assume the width  of the infinite well to be variable and expressed as
$D(y) = a + L(y) +w(y)$ with $a$ being a fixed constant. Inside the well we also assume {to add}
 a finite potential barrier of width $L(y)$ having constant hight $V_0:=\frac{9\pi^2\hbar^2}{2ma^2}$ and located at distance $a/2$ from 
 the left-most infinite wall,  see Fig.~\ref{fig:energies}(a). For $w(y)=0$,   $V_y(x)$ corresponds to a stretchable potential~\cite{inprep}  exhibiting a third energy level eigenvalue constantly equal to 
 $V_0$ which does not depend upon the selected value of $L(y)$. 
 \begin{figure}[!t]
 \centering
 \includegraphics[scale=0.35]{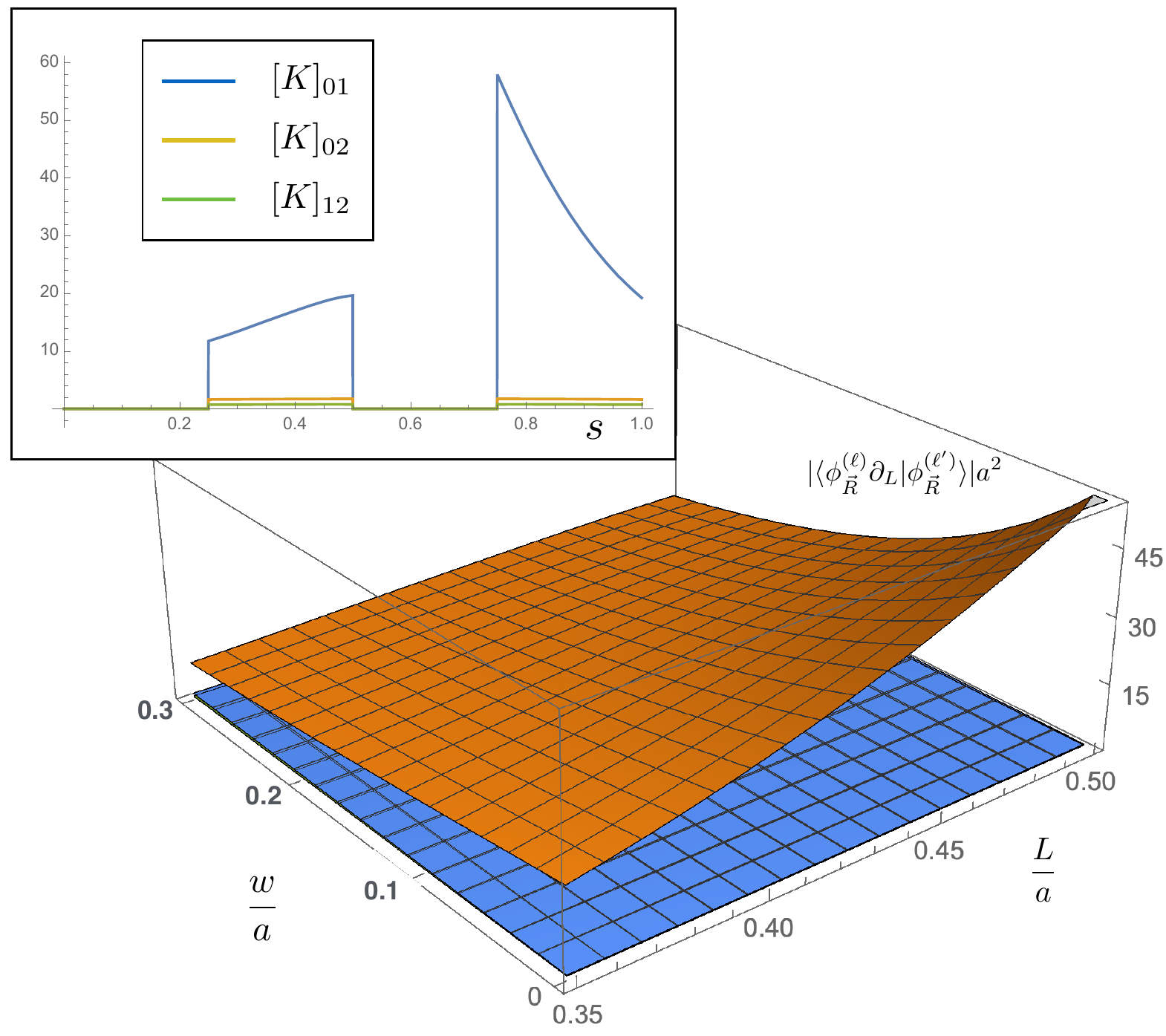}
 \caption{Plot of  $|\langle{\phi_{\vec{R}}^{(\ell)}}|{\partial_L\phi_{\vec{R}}^{(\ell')}}\rangle|a^2$ in the control
 plane $(L,w)$: $(\ell,\ell')=(0,1)$ (yellow surface) and $(1,2),(0,2)$ (blue surface, the differences between the two being not
 observable). 
Inset: Plot of the matrix elements $|K_{\ell\ell'}|$ when moving along a rectangular path ${\cal R}$   with $\vec{R}_{\text{in}}=(0.35a,0)$ and $\vec{R}_{\text{fin}}=(0.5a,0.02a)$-- see inset of 
 Fig.~\ref{fig:geo_phase}; $s$ being the length of the path along the trajectory expressed in unit of $a$. When $w$ is changed all the matrix elements are null, corresponding to the intervals $[0,1/4]$ and $[1/2,3/4]$, while as $L$ is changed (corresponding to the intervals $[1/4,1/2],[3/4,1]$) we can observe how $[K]_{01}$ is always far larger than $[K]_{12}$ and $[K]_{02}$.}
 \label{fig:matrix_elements}
 \end{figure}
 The energy landscape associated with the first three levels obtained by solving Eq.~(\ref{eq:schrodinger})
 is reported in Fig.~\ref{fig:energies} (b) as a function of the control parameters. 
We notice that as long as we prevent the ratio $w(y)/a$ to be above $\sim 0.05$  
the energy gap between the first two levels is much smaller than the one between these levels and the third, so that we are ensured that the matrix elements $[K_y]_{\ell\ell'}$ are negligible for $\ell,\ell'>2$ -- see Fig.~\ref{fig:matrix_elements}.  Moreover, in this region the energy gap between the ground state and the first excited level is also very small, ensuring that under WKB approximation the dynamical contribution to the system evolution 
will not add extra coupling terms that compete with the holomony. 
Therefore following the analysis of the previous section, we can safely consider the Hilbert space as two fold 
and compute the holonomy as in Eq.~(\ref{eq:holonomic_gate}).
\begin{figure}[!t]
\centering
\includegraphics[scale=0.5]{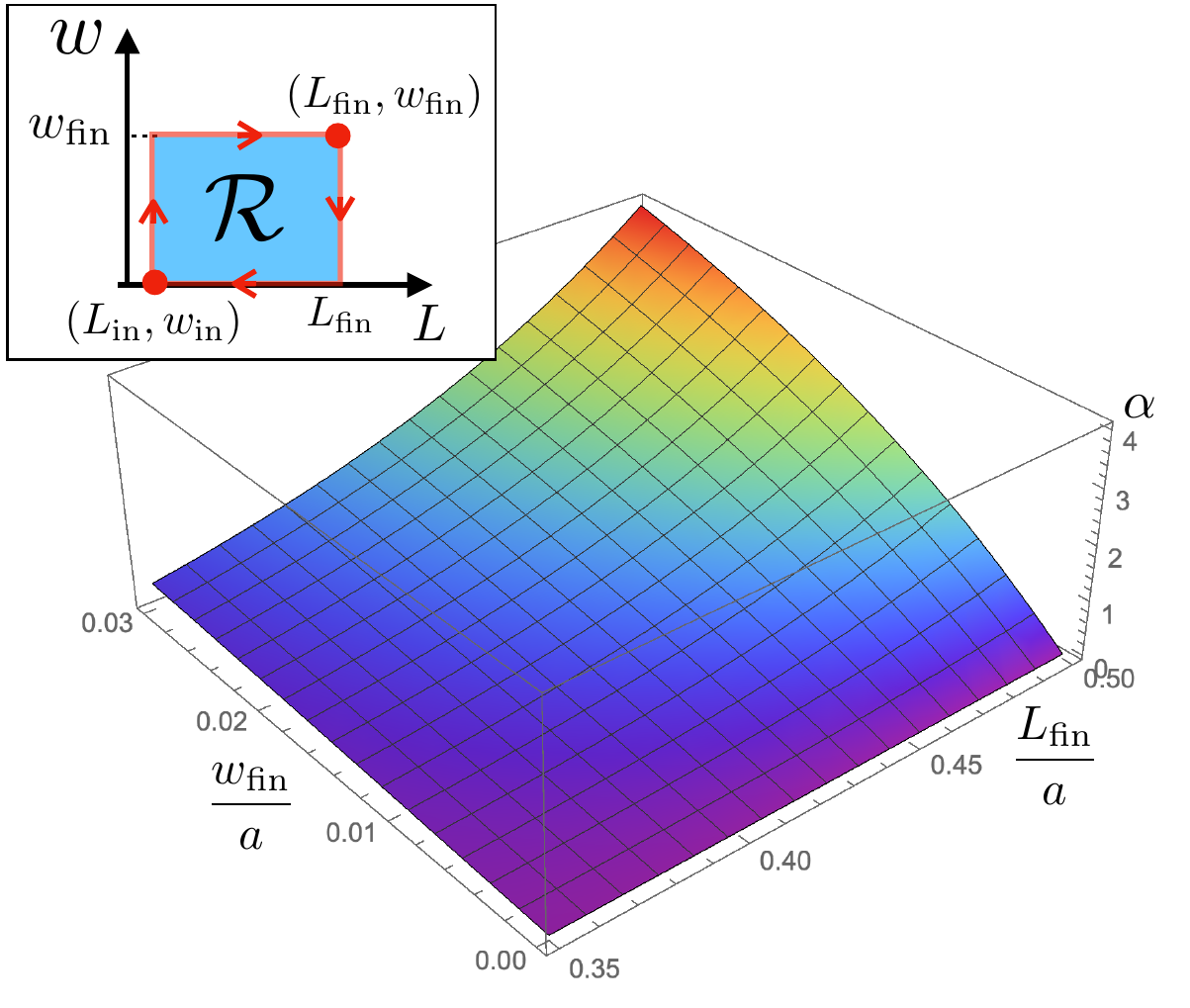}
\caption{Geometric phase~(\ref{DEFA})  obtained 
following a rectangular  closed path ${\cal R}$ in the  $(L,w)$-plane specified by the extremal points  
$\vec{R}_{\text{in}}=(L_{\text{in}},w_{\text{in}}) \rightarrow$ 
$(L_{\text{in}},w_{\text{fin}})$ $\rightarrow$  $(L_{\text{fin}},w_{\text{fin}})$ $\rightarrow$ $(L_{\text{fin}},w_{\text{in}})$ $\rightarrow$ $\vec{R}_{\text{in}}$, see inset. 
The plot refers to the case where we keep fixed $L_{\text{in}}=0.3a$ and $w_{\text{in}}=0$, while
spanning on $L_{\text{fin}}$ and $w_{\text{in}}$.}
\label{fig:geo_phase}
\end{figure}
We also observe that for the selected model, the second component of the vector $\vec{\lambda}(\vec{R})$ entering~(\ref{DEFA}) is always identically null, yielding 
$\vec{\lambda}({\vec{R}})=(\langle \phi^{(0)}_{\vec{R}}|\partial_L\phi^{(1)}_{\vec{R}}\rangle, 0)$.
Indeed we have 
 $\langle{\phi^{(0)}_{\vec{R}}}|{\partial_w\phi^{(1)}_{\vec{R}}}\rangle=0$ as it depends on the variation of the wave functions at the extremal point $x=a+L+w$ where the boundary conditions force both 
 the wave functions to be exactly null. Moreover, in Sec. VI of SM we have shown that the only non-zero component of 
 $\vec{\lambda}({\vec{R}})$ can be expressed as 
$\langle{\phi^{(0)}_{\vec{R}}}|{\partial_L\phi^{(1)}_{\vec{R}}}\rangle=-  \frac{9 \pi^2}{a^2}  \tfrac{\phi_{\vec{R}}^{*(0)}\left(\tfrac{a}{2}+L\right)\phi_{\vec{R}}^{(1)}\left(\tfrac{a}{2}+L\right)}{\epsilon^{(1)}_{\vec{R}} - \epsilon^{(0)}_{\vec{R}}}$,
where for $\ell=0,1$, $\phi_{\vec{R}}^{(\ell)}(x)$ and $\epsilon^{(\ell)}_{\vec{R}}$ represent
 the $\ell$-th eigenfunction and the associated (rescaled) eigenenergy of the Hamiltonian
 $\frac{\hat{p}^2_x}{2m}+{V}_{\vec{R}}(\hat{x})$.
Hence the  geometric phase~(\ref{DEFA}) computed 
 along the rectangular paths shown in the inset of Fig.~\ref{fig:geo_phase}
 can be expressed as 
$\alpha =\theta(w_{\text{fin}}) -  \theta(w_{\text{in}})$,
with $\theta(w)$ being the line integral defined as $\theta(w) :=  \frac{9 \pi^2}{a^2} \int_{L_{\text{in}}}^{L_{\text{fin}}} d L \;  {\phi_{\vec{R}}^{*(0)}\left(\tfrac{a}{2}+L\right)\phi_{\vec{R}}^{(1)}\left(\tfrac{a}{2}+L\right)}/({\epsilon^{(1)}_{\vec{R}} - \epsilon^{(0)}_{\vec{R}}})$.
As shown in Fig.~\ref{fig:geo_phase}  though moving in a small region of parameters space, we are able nonetheless to obtain a wide range of values of $\alpha$.
\paragraph{Conclusions and outlook:--}
Exploiting the spatial analogue of Berry phase
we have shown how it is possible through spatial potential engineering to induce a geometrical phase on the internal state of a traveling particle. In our analysis we assume that the kinetic energy 
of the particle is much larger than the relevant energy levels of the system and the energy scale associated with the potential modulation. This allows us to separate two main contributions in the solution of the time-independent Schr\"odinger equation: one which has a purely holonomic character, and a dynamical one which amounts to an irrelevant global phase.
The proposed scheme  might be envisioned as useful resource in the context of solid state devices quantum computing where potential profiles engineering 
has nowadays reached a sufficient level of precision needed for such applications. On the theoretical side it would be interesting to investigate how the presence of dissipation and particle interactions influence the appearance of a geometrical phase.

\begin{acknowledgments}
{\it Acknowledgments:--} S.C. and A.D.P.~contributed equally to this work. We would like to thank G. C. La Rocca for fruitful discussions and precious advice during the completion of this work, and R. Fazio e P.  Zanardi for their comments. S. C. would like to thank A. Carollo for fruitful discussions.
V. G. acknowledges support by MIUR via PRIN 2017 (Progetto di Ricerca di Interesse Nazionale): project QUSHIP (2017SRNBRK).
\end{acknowledgments}

\appendix
\section{SUPPLEMENTAL MATERIAL}
\setcounter{equation}{0}
\setcounter{figure}{0}
\setcounter{table}{0}
\setcounter{page}{1}
\renewcommand{\theequation}{S\arabic{equation}}
\renewcommand{\thefigure}{S\arabic{figure}}

The presented material is organized as follows:
In Sec.~\ref{SEC1} we discuss some basic properties of the matrix $K_y$ defined in Eq.~(4) of the main text.
In Sec.~\ref{SEC2} we give an explicit derivation of Eq.~(3) of the main text.
In Sec.~\ref{SEC3} we rewrite the holonomy operator (9) in terms of the curvature tensor of the model.
In Sec.~\ref{SECIV} we discuss the energy scale of the model and the WKB approximation~\cite{MESSIAH}. 
In Sec.~\ref{SECV} we give some technical details on the computation of the holonomy for the case of infinite
confining potetial.
In Sec.~\ref{SECVI} we present an example of a non trivial non-Abelian holonomy construction for three levels.

\section{Properties of the $K_y$ matrix} \label{SEC1} 
As anticipated in the main text the matrix $K_y$  is  explicitly anti-Hermitian, i.e. 
\begin{eqnarray} K_y^\dag = - K_y\label{ANTIHER} \;. \end{eqnarray} 
This property is fundamental for ensuring the unitarity of the associated holonomic transformation, i.e. the operator
\begin{widetext} 
\begin{eqnarray}&& {\cal U}_{Y_0\rightarrow y}=P
{\exp}\left[-\int^y_{{Y_0}} {K}_{y'}dy'\right]=\sum_{n=0}^\infty (-1)^n\int_{Y_0}^{y}dy_1
\int_{Y_0}^{y_1}dy_2\cdots  \int_{Y_0}^{y_{n-1}}dy_n K_{y_1}K_{y_2} \cdots K_{y_n}\;.
\end{eqnarray} 
\end{widetext}
which by construction fullfils the identity
\begin{eqnarray} \partial_y {\cal U}_{Y_0\rightarrow y} = -K_y  {\cal U}_{Y_0\rightarrow y}\;. 
\end{eqnarray} 
Equation~(\ref{ANTIHER})  is a direct consequence of the orthogonality conditions of the eigenvectors 
$\{ |{\phi_y^{(\ell)}}\rangle ; {\ell =0,1,2,\cdots}\}$, i.e. 
\begin{eqnarray} 
\langle \phi_y^{(\ell)}  |{\phi_y^{(\ell')}}\rangle = \delta_{\ell,\ell'} \;,
\end{eqnarray} 
which, upon differentiation with respect to $y$ gives
\begin{eqnarray} \label{PROPERTY1ofK} 
\nonumber
{[K_y]}_{\ell \ell'}&=&\langle \phi_y^{(\ell)}  |\partial_y {\phi_y^{(\ell')}}\rangle
=- \langle \partial_y \phi_y^{(\ell)}  |{\phi_y^{(\ell')}}\rangle\\&=&-\langle \phi_y^{(\ell')}  |\partial_y {\phi_y^{(\ell)}}\rangle^* = - {[K_y]}_{\ell' \ell}^* \;. 
\end{eqnarray} 
Notice also that thanks to the fact that $\{ |{\phi_y^{(\ell)}}\rangle ; {\ell =0,1,2,\cdots}\}$ form a complete
basis, we have

\begin{eqnarray} 
[ K_y^\dag K_y ]_{\ell\ell'}  &=& \sum_{\ell''} [ K_y^\dag]_{\ell \ell''} [K_y ]_{\ell'' \ell'} = 
\sum_{\ell''} [ K_y]_{\ell'' \ell}^* [K_y ]_{\ell'' \ell'}\nonumber\\
&=& \sum_{\ell''}    
\langle \phi_y^{(\ell'')}  |\partial_y {\phi_y^{(\ell)}}\rangle^* 
   \langle \phi_y^{(\ell'')}  |\partial_y {\phi_y^{(\ell')}}\rangle\nonumber\\&=&\sum_{\ell''}    
\langle\partial_y {\phi_y^{(\ell)}}|  \phi_y^{(\ell'')}  \rangle
   \langle \phi_y^{(\ell'')}  |\partial_y {\phi_y^{(\ell')}}\rangle\nonumber\\&=&  \langle\partial_y {\phi_y^{(\ell)}}|\partial_y {\phi_y^{(\ell')}}\rangle \;. \label{ffgd}
\end{eqnarray}

We can further observe that all the matrix elements ${[K_y]}_{\ell \ell'}$ (and hence the ${[J_y]}_{\ell \ell'}$ are real), i.e.
\begin{eqnarray} \label{PROPERTY2ofK} 
{[K_y]}_{\ell \ell'}  &=& {[K_y]}_{\ell \ell'}^* \;,
\end{eqnarray} 
which together with Eq.~(\ref{PROPERTY1ofK}) implies that its diagonal term are all null, i.e. 
\begin{eqnarray}  \label{IDO} 
{[K_y]}_{\ell \ell}  =0\;, \qquad \forall \ell. 
\end{eqnarray} 
The property (\ref{PROPERTY2ofK}) is 
 a consequence of the
fact that  $|{\phi_y^{(\ell)}}\rangle$ are eigensolutions of the 1D hamiltonian $\hat{h}_y(\hat{x}):=\frac{\hat{p}^2_x}{2m}+{V}_y (\hat{x})$. 
 In the $x$-representation the wave functions
 $\phi_y^{(\ell)}(x):=\langle x|\phi_y^{(\ell)} \rangle$
 associated with the eigenstates $|\phi_y^{(\ell)} \rangle$ can be always taken to be real together with all their derivative with respect to the parameter $y$, i.e. 
 \begin{eqnarray} \label{REAL} 
 \phi_y^{(\ell)}(x)  = {\phi_y^{(\ell)}}^*(x)\;,  \quad \partial_y^k  \phi_y^{(\ell)}(x) =(\partial_y^k  \phi_y^{(\ell)}(x))^* \;.
 \end{eqnarray} 
 From this it then follows that 
 \begin{widetext}
\begin{eqnarray} \label{IMPOIMPO} 
 &&\langle \phi_{y'}^{(\ell')}   |\partial_y{\phi}_y^{(\ell)} \rangle = \int dx\;  {\phi_{y'}^{(\ell')}}^*(x) \partial_y{\phi}_y^{(\ell)}(x)= \int dx \phi_{y'}^{(\ell')}(x)(\partial{\phi}_y^{(\ell)}(x))^* =
\langle \partial{\phi}_y^{(\ell)}| \phi_{y'}^{(\ell')}   \rangle=\langle 
\phi_{y'}^{(\ell')} |    \partial{\phi}_y^{(\ell)}\rangle^*\;, \end{eqnarray} \end{widetext}
 which establishes that these quantities must be real for all $\ell,\ell', y$ and $y'$ --  
 Eq.~(\ref{PROPERTY2ofK}) finally follows from this by simply setting $y'=y$. 

Exploiting the fact that $|\phi_y^{(\ell)}\rangle$ is eigenvector of $\hat{h}_y(\hat{x})$ 
with eigenvalue $E_y^{(\ell)}$ we can write 
\begin{eqnarray} 
\label{eq:identity_2}
\langle{\phi_{y'}^{(\ell')}}|\hat{h}_y(\hat{x})|{\phi_y^{(\ell)}}\rangle=E_y^{\ell} \langle{\phi_{y'}^{(\ell')}}|{\phi_y^{(\ell)}}\rangle.
\end{eqnarray} 
By deriving this expression with respect to $y$ and then setting $y'=y$ this finally leads to 
\begin{eqnarray} 
{[K_y]}_{\ell' \ell} &=& \langle{\phi_{y}^{(\ell')}}|{\partial_y{\phi}_y^{(\ell)}}\rangle=\frac{\langle{\phi_y^{(\ell')}}|\left( \partial_y{\hat{h}}_y(\hat{x})\right) |{\phi_y^{(\ell)}}
\rangle}{E_y^{(\ell)}-E_y^{(\ell')}}\nonumber\\&=&\frac{\langle{\phi_y^{(\ell')}}|\left( \partial_y{{V}}_y(\hat{x})\right) |{\phi_y^{(\ell)}}\rangle}{E_y^{(\ell)}-E_y^{(\ell')}},
\end{eqnarray} 
which holds for all $\ell\neq \ell'$ and which using Eq.~(\ref{IDO}) finally gives us Eq.~(4) of main text. 

\section{Derivation of Eq.~(3)} \label{SEC2} 
Given $\ket{\psi_E(y)}$ solution of Eq.~(1) of the main text, 
we can expand it in terms of 
the orthonormal set $\{ |{\phi_y^{(\ell)}}\rangle ; {\ell =0,1,2,\cdots}\}$ formed by the egeinsolutions of the
1D Hamiltonian $\hat{h}_y(\hat{x}):=\frac{\hat{p}^2_x}{2m}+{V}_y (\hat{x})$, i.e. 
is defined by the solution 
\begin{eqnarray}  \label{expansion} 
\ket{\psi_E(y)}&=&\sum_\ell C_y^{(\ell)}|{\phi_y^{(\ell)}}\rangle \;,
\end{eqnarray} 
with $C_y^{(\ell)} :=  \langle{\phi_y^{(\ell)}}  \ket{\psi_E(y)}$ and 
\begin{eqnarray} 
\hat{h}_y(\hat{x})|{\phi_y^{(\ell)}}\rangle &=& E_y^{(\ell)} |{\phi_y^{(\ell)}}\rangle  \;.\end{eqnarray} 
Observe then that\begin{widetext}
 \begin{eqnarray} 
 \nonumber
 \langle{\phi_y^{(\ell)}}  \ket{\partial_y^2\psi_E(y)} =   \partial^2_y C_y^{(\ell)}+ 
\sum_{\ell'} 2 \partial_y C_y^{(\ell')} \langle{\phi_y^{(\ell)}}   |\partial_y {\phi_y^{(\ell')}}\rangle+\sum_{\ell'} C_y^{(\ell')} \langle{\phi_y^{(\ell)}}   |\partial^2_y {\phi_y^{(\ell')}}\rangle
=  \partial^2_y C_y^{(\ell)} + \sum_{\ell'} \left(  2 [ K_y ]_{\ell \ell'} \partial_y C_y^{(\ell')}  
+  [ \Gamma_y ]_{\ell \ell'} C_y^{(\ell')}\right)  \;,
 \end{eqnarray} \end{widetext}
 with $K_y$ and $\Gamma_y$ the matrices of elements
 \begin{equation} 
 {[K_y]}_{\ell \ell'}  := \langle \phi_y^{(\ell)}  |\partial_y {\phi_y^{(\ell')}}\rangle\;,  \qquad 
 {[ \Gamma_y ]}_{\ell \ell'}  :=  \langle{\phi_y^{(\ell)}}   |\partial^2_y {\phi_y^{(\ell')}}\rangle  \;, 
 \end{equation} 
 Replacing this into Eq.~(1) of the main text gives us the following set of equations for the coefficient  $C_y^{(\ell)}$ 
\begin{eqnarray} 
  \partial^2_y C_y^{(\ell)} &+&\sum_{\ell'} \Big(  2 [ K_y ]_{\ell \ell'} \partial_y C_y^{(\ell')} \nonumber\\
&+&  [ \Gamma_y ]_{\ell \ell'} C_y^{(\ell')}\Big)+\tfrac{2m }{\hbar^2} (E- E_y^{(\ell)}) C_y^{(\ell)} =0 \;\,
\end{eqnarray}
or, in vectorial form
 \begin{eqnarray} \label{rima} 
  \partial^2_y {{\bf C}_y}   &+& 2 K_y  \partial_y {{\bf C}_y}  + (\Gamma_y + \epsilon - \Omega_y)  {{\bf C}_y} 
 =0 \;, 
\end{eqnarray} 
with $\epsilon= \tfrac{2m }{\hbar^2} E$ and with 
 $\Omega_y$ the matrix of elements
\begin{eqnarray} 
{[\Omega_y]}_{\ell \ell'} = \tfrac{2m }{\hbar^2}  E_y^{(\ell)} \delta_{\ell \ell'} \;. \end{eqnarray} 
Equation (\ref{rima}) can finally casted in the form in Eq.~(3) 
by taking
\begin{eqnarray}
 \Gamma_y  - \partial_y K_y - K_y^2  =0\;,
\end{eqnarray} 
that follows   from the fact that  
\begin{eqnarray} 
\nonumber
[\Gamma_y - \partial_y K_y]_{\ell\ell'}&=& \langle{\phi_y^{(\ell)}}   |\partial^2_y {\phi_y^{(\ell')}}\rangle - \partial_y\left( \langle \phi_y^{(\ell)}  |\partial_y {\phi_y^{(\ell')}}\rangle\right)\\&=&-
 \langle{\partial_y \phi_y^{(\ell)}}   |\partial_y {\phi_y^{(\ell')}}\rangle = [ K_y^2 ]_{\ell\ell'}\;,
\end{eqnarray} 
where we used (\ref{ffgd}) and  from Eq.~(\ref{ANTIHER}) which implies
$K_y^2 = - K_y^\dag K_y$.

\section{Energy scales analysis and  WKB approximation} \label{SECIV} 
The starting assumption of our analysis is that  the modulations of the transverse confining potential ${V}_y(\hat{x})$
is such that the matrix $K_y$ of Eq.~(4) of the main text only couples energy levels which are sufficiently close in energy. 
In particular we shall assume that there exist $\ell_{\rm off} \geq \ell_0$ such that \begin{widetext}
\begin{eqnarray}   \int_{Y_0}^Y dy \; \Big|  \frac{ \langle{\phi_{y}^{(\ell)}}|   \partial_y{{V}}_y(\hat{x}) |{\phi}_y^{(\ell')}  \rangle}
 {E_y^{(\ell')}- E_y^{(\ell)} } \Big| = 
\tfrac{2m}{\hbar^2  }\ \int_{Y_0}^Y dy \; \Big|  \frac{ \langle{\phi_{y}^{(\ell)}}|   \partial_y{{V}}_y(\hat{x}) |{\phi}_y^{(\ell')}  \rangle}
 {\epsilon_y^{(\ell')}- \epsilon_y^{(\ell)} } \Big| \ll 1\;, \label{cutoff} \qquad \forall \ell' \leq \ell_0\;, \quad   \forall \ell \geq  \ell_{\rm off}+1\;, 
\end{eqnarray} \end{widetext}
which can be interpreted as an adiabatic condition for the  modulation of the potential.
Accordingly if the particle enters the scattering region with input states that have a non zero overlap with only  the first $\ell_0$ 
low energy levels (i.e. ${\bf C}_{Y_0} \simeq  (C_{Y_0}^{(0)}, C_{Y_0}^{(1)},\cdots,
C_{Y_0}^{(\ell_0)}, 0,\cdots, 0)$), 
we can focus  on solutions of Eq.~(5) of the form 
 $\tilde{\bf C}_y \simeq  (C_y^{(0)}, C_y^{(1)},\cdots,
C_y^{(\ell_{\rm off})}, 0,\cdots, 0)$ which only involve the levels up to the
cut-off threshold  $\ell_{\rm off}$ set in Eq.~(\ref{cutoff}), or equivalently
 identify  all the matrices entering in Eq.~(5) with their
  $\ell_{\rm off}\times \ell_{\rm off}$ principal minors associated with  the first $\ell_{\rm off}$ levels. 
Of course, to avoid trivial results we need also to ensure that when evaluated on the relevant energy levels 
the right-hand-side quantity of Eq.~(\ref{cutoff}) will not be negligible, a condition that,
 when 
selection rules that explicitly impose $\langle{\phi_{y}^{(\ell)}}|   \partial_y{{V}}_y(\hat{x}) |{\phi}_y^{(\ell')}  \rangle=0$ for $\ell, \ell' \leq \ell_0$ do not hold,
can always be enforced under the quasi-degenerate condition
\begin{eqnarray}\label{degcOn} 
\epsilon_y^{(\ell_0)} -\epsilon_y^{(0)} \ll \epsilon_y^{(\ell_{\rm off}+1)} -\epsilon_y^{(\ell_0)}\;.\end{eqnarray}

Our second major assumption is that the kinetic energy of the particle sets the largest energy scale of the system, a condition which in view of the above considerations translates
into the inequality 
\begin{eqnarray} \epsilon \gg  \epsilon_y^{(\ell)} \;, \label{cond1234} 
\end{eqnarray} 
 for all $y$ in the scattering domain and for all the energy levels  which are involved in the solution of the
dynamical equation (5), i.e. 
$\ell\leq \ell_{\rm off}$, or equivalently into the condition 
\begin{eqnarray} \epsilon \gg  \|\Omega(y)\|  = \|\tilde{\Omega}(y)\|  \;, \label{COND11112}
\end{eqnarray} 
the last identity following from the fact that $\Omega(y)$ and $\tilde{\Omega}(y)$ are equivalent under the unitary
transformation 
 ${\cal U}_{Y_0\rightarrow y}=P
{\exp}\left[-\int^y_{{Y_0}} {K}_{y'}dy'\right]$.
Thanks to Eq.~(\ref{COND11112}) we can now  rewrite Eq.~(5) as 
\begin{equation}\partial_y ^2 \tilde{\bf C}_y   = -\tilde{\kappa}_y^2 \tilde{\bf C}_y \;, 
 \label{eq:differential_2SUPP}
\end{equation} 
where $\tilde{\kappa}_y$ is  the  positive operator \begin{eqnarray}  \label{DEFKAPPA} 
&&\tilde{\kappa}_y : = \sqrt{\epsilon - \tilde{\Omega}(y)}={\cal U}_{Y_0\rightarrow y}^\dag \;
 \kappa_y \;   {\cal U}_{Y_0\rightarrow y} \;,\\ \nonumber &&\kappa_y:=  \sqrt{\epsilon - {\Omega}(y)}
\;,\end{eqnarray} 
represents the effective spatially-dependent wave-vectors describing the longitudinal propagation of the particle along the sample. 
For $\tilde{\kappa}_y=\tilde{\kappa}$ constant (a condition which holds true when the confining potential of the model is  invariant  under longitudinal translations $\partial_y{V}_y(\hat{x})=0$), Eq.~(\ref{eq:differential_2SUPP}) admits the simple plane-wave solution
\begin{eqnarray} \tilde{\bf{C}}_{y}= e^{ i \tilde{\kappa} (y- Y_0)} {\bf{A}} +  e^{- i\tilde{\kappa} (y- Y_0)}  {\bf{B}} \;,\end{eqnarray} 
with ${\bf{A}}$ and ${\bf{B}}$ fixed by the boundary conditions of the problem. An analogous expression applies also if 
$\kappa_y$ is a sufficiently smooth functions of the coordinate $y$. Specifically  imposing the condition 
\begin{eqnarray} \label{condWKB} 
\| \partial_y \tilde{\kappa}_y\| \ll  \|\tilde{\kappa}_y\|^2\;, 
\end{eqnarray} 
we can invoke  WKB approximation~\cite{MESSIAH} to write the solution of Eq.~(\ref{eq:differential_2SUPP}) (and hence of 
Eq.~(5) of the main text) as 
\begin{eqnarray} \tilde{\bf{C}}_{y}\simeq {\cal W}^{(+)}_{y}   {\bf{A}} +  {\cal W}^{(-)}_{y}   {\bf{B}} \label{WKB} \;,\end{eqnarray} 
where now ${\cal W}^{(\pm)}_{y}$ are the path-ordered exponentials
\begin{widetext}
\begin{eqnarray} {\cal W}_{y}^{(\pm)} : =P
{\exp}\left[\pm i \int^y_{{Y_0}} \tilde{\kappa}_{y'} dy'\right]=\sum_{n=0}^\infty (\pm i )^n\int_{Y_0}^{y}dy_1
\int_{Y_0}^{y_1}dy_2\cdots  \int_{Y_0}^{y_{n-1}}dy_n \tilde{\kappa}_{y_1}\tilde{\kappa}_{y_2} \cdots \tilde{\kappa}_{y_n}\;,
\end{eqnarray}
\end{widetext}
which, as anticipated in the main text, at the lowest order in (\ref{COND11112})
can be estimated as  the multiplicative  phases ${\cal W}_{y}^{(\pm)} \simeq e^{ \pm i \sqrt{\epsilon} (y- Y_0)}$,
or as  ${\cal W}_{Y}^{(\pm)}  \simeq e^{\pm i \int_{Y_0}^Y \sqrt{\epsilon  - \omega_{y'}} dy'}$ as reported in the  {\it Two-dimensional models} section.  
\\

Let us now elaborate a bit on Eq.~(\ref{condWKB}).
Notice first that from Eq.~(\ref{DEFKAPPA}) we get 
\begin{eqnarray} 
&&\partial_y \tilde{\kappa}_y = {\cal U}_{Y_0\rightarrow y}^\dag \left(\partial_y \kappa_y -  [ \kappa_y,K_y]  \right) {\cal U}_{Y_0\rightarrow y} \;,\\\nonumber
&&\partial_y {\kappa}_y =-  \frac{\partial_y \Omega_y}{ 2{\kappa_y}} \;, 
\end{eqnarray} 
where in writing the second identity we explicitly use the fact that $\Omega_y$ (and hence $\kappa_y$ and $\partial_y \Omega_y$)
are diagonal matrices. 
This allows us to recast~(\ref{condWKB}) into the equivalent form 
\begin{eqnarray} \label{condWKB_1} 
\Big\|  \frac{\partial_y \Omega_y}{ {\kappa_y}} + 2   [ \kappa_y,K_y] 
\Big\| \ll  2 \|{\kappa}_y\|^2\;.
\end{eqnarray} 
Notice that, while the right-hand-side part of the above expression does not depend upon the first derivative of 
$\hat{V}_y(\hat{x})$ with respect to the longitudinal coordinate $y$, 
 both terms entering in the left-hand-side of the above expression do. In the case of 
$[ \kappa_y,K_y] $ this is directly evident from Eq.~(4). For $\partial_y \Omega_y$ this can be easily verified by observing that 
 $[\Omega_y]_{\ell \ell'}  = \epsilon_y^{(\ell)} \delta_{\ell\ell'}$ is the diagonal matrix of elements given by the (rescaled) eigenvalues $\epsilon_y^{(\ell)}$ of the operator $\hat{h}_y(\hat{x})=\frac{\hat{p}^2_x}{2m}+{V}_y (\hat{x})$.
Writing then $\epsilon_y^{(\ell)} =  \tfrac{2m}{\hbar^2  }  \langle {\phi_y^{(\ell)}}| \hat{h}_y(\hat{x})|{\phi_y^{(\ell)}}\rangle$ and using 
Eq.~(\ref{PROPERTY1ofK}) it follows that 
\begin{eqnarray} 
\nonumber
 \partial_y \epsilon_y^{(\ell)}  &=&    \tfrac{2m}{\hbar^2  }    \langle {\phi_y^{(\ell)}}| \left(  \partial_y   \hat{h}_y(\hat{x})\right) |{\phi_y^{(\ell)}}\rangle\\&=&
   \tfrac{2m}{\hbar^2  }    \langle {\phi_y^{(\ell)}}| \left(  \partial_y   \hat{V}_y(\hat{x})\right) |{\phi_y^{(\ell)}}\rangle \;, 
\end{eqnarray} 
from which we have 
\begin{eqnarray} 
[{\partial_y \Omega_y}]_{\ell \ell'} =    \tfrac{2m}{\hbar^2  }   \delta_{\ell\ell'}  \;     \langle {\phi_y^{(\ell)}}| \left(  \partial_y   \hat{V}_y(\hat{x})\right)|{\phi_y^{(\ell)}}\rangle \;. 
\end{eqnarray} 
To translate Eq.~(\ref{condWKB_1}) in a more quantitive condition let us recall (\ref{COND11112})  to approximate $\kappa_y\simeq \sqrt{\epsilon}$ and 
$[ K_y,\kappa_y] \simeq 0$ obtaining $\| {\partial_y \Omega_y}\| \ll  2 \epsilon^{3/2}$ or
\begin{eqnarray} \label{condWKB1212} 
    |\langle {\phi_y^{(\ell)}}| \left(  \partial_y   \hat{V}_y(\hat{x})\right)|{\phi_y^{(\ell)}}\rangle|  \ll \frac{\hbar^2  \epsilon^{3/2}}{m} =  \sqrt{ \frac{8m E^3}{\hbar^2} } \;,\, 
\end{eqnarray} 
for all $y$ in the domain, and for all $\ell\leq \ell_{\rm off}$. 

\section{Curvature tensor formula} \label{SEC3}
The exponential operator ${\cal U}({\cal R})$  appearing in Eq.~(9) of the main text is formally defined as \begin{widetext}
  \begin{eqnarray}
 {\cal U}({\cal R})= P \exp[ - \oint_{\cal R}  d \vec{R} \cdot \vec{{K}}(\vec{R})]=\sum_{n=0}^\infty (-1)^n\int_{\vec{R}_0}^{\vec{R}_Y}d\vec{r}_1\cdot {\vec{K}}(\vec{r}_1)
\int_{\vec{R}_0}^{\vec{r}_{1}}d\vec{r}_2\cdot {\vec{K}}(\vec{r}_2)\cdots \int_{\vec{R}_0}^{\vec{r}_{n-1}}d\vec{r}_n\cdot {\vec{K}}(\vec{r}_n)\;,
\end{eqnarray}\end{widetext}
where given  $j=0,\cdots, {n-1}$,   $\vec{r}_j:= \vec{R}_{y_j}$ is the element 
of the curve ${\cal R}$ assumed by the control vector $\vec{R}_y$ at the  point $y=y_j$, with 
 $y_1,y_2, \cdots y_n$ being coordinate values in $[Y_0,Y[$ which are ordered so that $y_{j}\geq y_{j+1}$.
 Following~\cite{HALPEN} 
it can be casted in the form 
\begin{eqnarray} 
 {\cal U}({\cal R})= {\cal P} \exp[\frac{i}{2} \int_{\cal S} {\cal F}_{i,j}(\vec{R}) \; dR^{(i)} \wedge dR^{(j)} ]\;,\end{eqnarray}   
where ${\cal S}$ is any regular surface in the control space which admits $\cal R$ as bounding curve, 
where the symbol ${\cal P}$ remind us that the integral must be performed under surface-ordering, 
and where finally 
  \begin{equation} {\cal F}_{i,j}(\vec{R}):= i \tfrac{\partial {K}^{(j)}(\vec{R})}{\partial R^{(i)}}-i \tfrac{\partial { K}^{(i)}(\vec{R})}{\partial R^{(j)}}+ i
\left[{ K}^{(i)}(\vec{R}),{ K}^{(j)}(\vec{R})\right]\;,\end{equation} 
is the Berry curvature tensor associated with the vector field $\vec{{K}}(\vec{R})$, i.e. 
see~\cite{STOKES1} and references therein. 

\section{Example of non-Abelian holonomy}\label{SECVI} 
\begin{figure}[!t]
\centering
\includegraphics[scale=0.35]{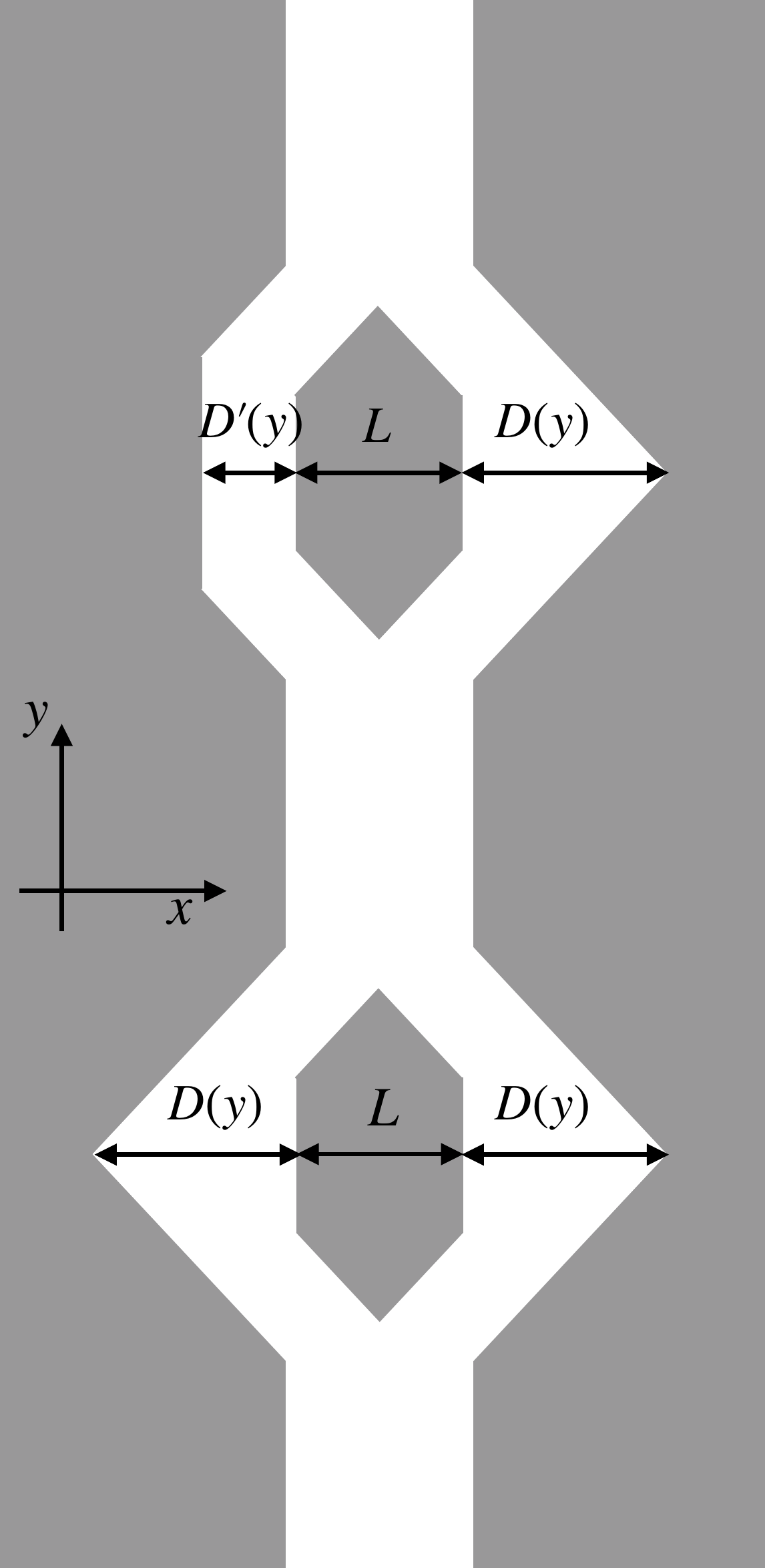}
\caption{Example of two geometric paths generating an holonomy: white and grey regions represent respectively the areas where the potential is zero and infinite, respectively. A particle entering  the potential landscape from the bottom will first undergo a unitary transformation involving $|\phi_y^{(0)}\rangle$ and $|\phi_y^{(2)}\rangle$, since symmetry constraints imply zero coupling between the groundstate and the first excited state (and also between $|\phi_y^{(1)}\rangle$ and $|\phi_y^{(2)}\rangle$). Then, the particle goes through an asymmetric landscape potential, inducing a gate which involves only levels $|\phi_y^{(0)}\rangle$ and $|\phi_y^{(1)}\rangle$. Therefore, since the generators of two transformations do not commute, the overall gate induces is a non-abelian holonomy.}
\label{example}
\end{figure}

Herewith we provide an example of  how to generate an holonomy within a subgroup of $SU(3)$ exploiting
a proper concatenation of non-commuting transformations. To this end let us consider the setting presented in Fig.~\ref{example} which exhibits two independent regions of modulation of the  control parameters space. 
In our analysis we shall assume that in this scenario, the conditions (\ref{cutoff}), (\ref{degcOn}), (\ref{cond1234}), and (\ref{condWKB1212})
detailed in Sec.~\ref{SECIV} of the Supplemental Material
hold true having  set $\ell_0=\ell_{\rm off}=2$, i.e. having identified the relevant energy levels with the three lowest ones.

The first modulation region from the bottom is symmetric along the $x$-axis (with respect to the $y$-axis) thus implying that the elements 
of the matrix $[K_y]_{\ell\ell'}$ of Eq.~(4) of the main text that connect  states $|{\phi_y^{(\ell)}}\rangle$,
$|{\phi_y^{(\ell')}}\rangle$ with different parities (e.g. $(\ell,\ell')=(0,1)$ or $(1,2)$) are null. This guarantees that the associated holonomy
 will involve only the even energy levels $(\ell,\ell')=(0,2)$
 with a contribution of the form
 $e^{i\alpha'{\sigma}_2^{(0,2)}}$ where ${\sigma}_2^{(\ell,\ell')}$  indicates the second Pauli operator acting on levels $(\ell,\ell')$ and $\alpha'$ being computed as in Eq.~(13) of the main text.
  Next, the particle enters the second modulation region which is asymmetric along $x$ and fulfils the same conditions we fixed in the section~{\it Two-dimensional models} of the main text
  therefore coupling the groundstate $|\phi_y^{(0)}\rangle$ with the first excited level $|\phi_y^{(1)}\rangle$ 
  via the gate defined in Eq.~(11) of the main text. The overall transformation will be therefore given by:
\begin{eqnarray} \label{dde1} 
\mathcal{U}_{Y_0\rightarrow Y}=e^{i\alpha {\sigma}_2^{(0,1)}}e^{i\alpha' {\sigma}_2^{(0,2)}}\;.
\end{eqnarray}
Since ${\sigma}_2^{(0,1)}$ and ${\sigma}_2^{(0,1)}$ do not commute, the holonomy described in Eq.~(\ref{dde1})  
generate a non-Abelian  subgroup of $SU(3)$.

\section{Details on the Holomomy calculation for the structured, infinite potential well} \label{SECV} 
Following the same derivation that leads to Eq.~(4) 
we can express the matrix $[\vec{{K}}(\vec{R})]_{\ell\ell'}$ of Eq.~(8) as 
\begin{equation} \label{IMPOKYs} 
\langle{\phi_{\vec{R}}^{(\ell)}}|{\vec{\nabla}_{\vec{R}} {\phi}_{\vec{R}}^{(\ell')}}\rangle
=
\frac{\langle{\phi_{\vec{R}}^{(\ell)}}|\big( \vec{\nabla}_{\vec{R}} {{V}}_{\vec{R}}(\hat{x})\big) |{\phi_{\vec{R}}^{(\ell')}}\rangle}{E_{\vec{R}}^{(\ell')}-E_{\vec{R}}^{(\ell)}} (1- \delta_{\ell\ell'})\;.
\end{equation} 
where for $\ell=0,1$, $\phi_{\vec{R}}^{(\ell)}(x)$ and $\epsilon^{(\ell)}_{\vec{R}}$ represent
 the $\ell$-th eigenfunction and the associated (rescaled) eigen-energy of the Hamiltonian
 $\frac{\hat{p}^2_x}{2m}+{V}_{\vec{R}}(\hat{x})$, the vector ${\vec{R}}$ encoding as usual the full
 dependence upon the longitudinal axis $y$. 
To proceed with the analysis we find it useful to replace the structured, infinite potential well of Fig.~2 (a)
with a regularized version that lives on the whole real axis. Specifically we consider the potential 
$V^{(\kappa)}_{y}(x)= V^{(\kappa)}_{\vec{R}_y}(x)$ where, indicating with 
$L$ and $w$ the cartesian components of the control vector
 $\vec{R} = (L,w)$, we define \begin{widetext}
\begin{eqnarray} 
&&V^{(\kappa)}_{\vec{R}}(x) := V_0 \Big[ \Theta(x-a/2) - \Theta(a/2+ L -x)  + \kappa  f(x)  \Theta(-x) +\kappa  f(x- a- L -w) \Theta(x- a- L -w) \Big]\;.
\end{eqnarray} \end{widetext}
In the above expression $\Theta(x)$ is the Heaviside step function, where $f(x)$ is 
any  regular function that nullifies in $x=0$ and strictly positive everywhere else (e.g. 
$f(x)=1-e^{-(x/a)^4}$), 
and where finally 
$\kappa\gg1$ is the regularization parameter which in the end will be sent to infinity to recover
$V_{\vec{R}}(x)$.
\begin{widetext}
Notice then that for the following identities hold
\begin{eqnarray} 
\partial_L V^{(\kappa)}_{\vec{R}}(x) &=& - V_0  \delta(a/2+ L -x)-  \kappa V_0 f(x- a- L -w)  \delta(x- a- L -w)= - V_0  \delta(a/2+ L -x)\;, \\ 
\partial_w V^{(\kappa)}_{\vec{R}}(x) &=& - \kappa V_0 f(x- a- L -w)  \delta(x- a- L -w)=0\;.
\end{eqnarray} 
Therefore, for all finite $k$  we can write
\begin{eqnarray} \label{IMPOKYs1} 
\langle{\phi_{{\vec{R}}}^{(\ell,\kappa)}}| (\partial_L V^{(\kappa)}_{\vec{R}}(x))| {\phi}_y^{(\ell',\kappa)}\rangle&=& 
   - V_0
 \phi_{\vec{R}}^{*(\ell,\kappa)}\left(\tfrac{a}{2}+L \right) \phi_{\vec{R}}^{(\ell',\kappa)}\left(\tfrac{a}{2}+L\right) 
 \;, \\ 
 \langle{\phi_{y}^{(\ell,\kappa)}}| (\partial_w V^{(\kappa)}_y(x))| {\phi}_y^{(\ell',\kappa)}\rangle&=& 0 \;,
\end{eqnarray} 
with $|{\phi}_{\vec{R}}^{(\ell,\kappa)}\rangle$ being the $\ell$-th energy eigenvector of the 
Hamiltonian associated with the $\kappa$-regularized potential.
Taking the $\kappa\rightarrow\infty$ limit and using Eq.~(\ref{IMPOKYs}) 
 this finally yields 
\begin{eqnarray} \label{IKM}
\langle{\phi^{(\ell)}_{\vec{R}}}|{\partial_L\phi^{(\ell')}_{\vec{R}}}\rangle &=&  \tfrac{-V_0}{E^{(\ell')}_{\vec{R}} - E^{(\ell)}_{\vec{R}}}  
\phi_{\vec{R}}^{*(\ell)}\left(\tfrac{a}{2}+L\right)\phi_{\vec{R}}^{(\ell')}\left(\tfrac{a}{2}+L\right)(1- \delta_{\ell\ell'})\;,  \\ 
\langle{\phi^{(\ell)}_{\vec{R}}}|{\partial_w\phi^{(\ell')}_{\vec{R}}}\rangle &=& 0\;.
\end{eqnarray} \end{widetext}
Notice in particular that taking $\ell=0$ and $\ell'=1$, Eq.~(\ref{IKM}) 
 reduces to the expression reported in the main text once we express the energies in terms of their
 rescaled counterparts $\epsilon^{(\ell)}_{\vec{R}}:= \tfrac{2m }{\hbar^2}E^{(\ell)}_{\vec{R}}$
and when we use the fact that $V_0=\frac{9\pi^2\hbar^2}{2ma^2}$.

\end{document}